\documentclass[aps,pra,twocolumn,amsmath,amssymb,showpacs]{revtex4}
\usepackage{graphicx}
\usepackage{dcolumn}
\usepackage{bm}
\usepackage{amsmath}
\input epsf

\begin{document}

\preprint{}

\title{Product state control of bi-alkali chemical reactions}

\author{Edmund R. Meyer}
\email{meyere@murphy.colorado.edu}
\affiliation{JILA, NIST and University of Colorado, Department of Physics,
  Boulder, Colorado 80309-0440, USA}
\author{John L. Bohn}
\affiliation{JILA, NIST and University of Colorado, Department of Physics,
  Boulder, Colorado 80309-0440, USA}

\date{\today}

\begin{abstract}
We consider ultracold, chemically  reactive scattering collisions of 
the diatomic molecules KRb.  When two such molecules collide in an
ultracold gas, we find that
they are energetically forbidden from reacting to form the trimer species
K$_2$Rb or Rb$_2$K, hence can only react via the bond-swapping reaction
2KRb $\rightarrow$ K$_2$ $+$ Rb$_2$.  Moreover, the tiny energy released in this
reaction can in principle be set to zero by applying electric or microwave 
fields, implying a means of controlling the available reaction channels in 
a chemical reaction.
\end{abstract}


\maketitle

The business of chemical physics is to understand the transformation of
reactant molecules into product molecules during a chemical reaction.
On the theory side, this is a daunting task, requiring the construction of
elaborate multidimensional potential energy surfaces, complemented by
classical, semiclassical, or even fully quantum mechanical scattering
calculations on these surfaces.  For experiments, the goal is to
provide as complete a picture as possible via the complete
determination of all initial and final states.  For initial states,
the use of molecular beam techniques provides excellent selection of
internal degrees of freedom in the reactant molecules, and control
over their relative state of motion.  For final states, spectroscopic
methods can select the relative abundance of the different rotational
and vibrational states of the products, providing a wealth of
information from which reaction dynamics can be inferred \cite{Levine}.

Recently, the experimental attainment of ultracold molecules has
pushed molecular beam technology to its ultimate limit.  It is
now possible to prepare a molecule in a single quantum state in all
degrees of freedom, down to the nuclear spin 
state~\cite{Ni08_Science,Danzl09_preprint}.  The molecules are
moreover characterized by  an extremely narrow range of
velocities, as set by gaseous temperatures on the order of $\mu $K.
This circumstance has enabled orders-of-magnitude control over
chemical reaction rates, by simply altering the nuclear spin 
state of the reactant molecules \cite{Ospelkaus10_Science}
or else by applying an electric field \cite{Ni10_Nature_preprint}.  
To probe reactions further, 
one could imagine transferring the molecules from the prepared 
ground state to any ro-vibrational excited state of interest.
These advances suggest what is possible when one prepares and
manipulates the {\it initial} states of the chemical reaction.

In this Letter, we suggest that manipulation of the {\it final} states
may also be possible in certain circumstances.  Specifically,
exit channels can become either energetically allowed, or else
energetically disallowed, as a function of electric field ${\cal E}$.
This idea was advanced previously, in the context of the H+LiF $\rightarrow$
Li + HF reaction, where it was argued that the exothermicity of
the reaction could be shifted via electric fields~\cite{Tscherbul08_JCP}.
Here we point out
that, for alkali dimers at ultralow temperature, it is conceivable
that a previously exothermic reaction can be completely turned off.
This possibility is afforded
by the fact that, for alkali dimers, the final states of reaction
are not terribly different in energy from the reactants.  Specifically,
we consider collisions of a pair of KRb molecules, prepared in
particular rotational states $n_1$ and $n_2$. These molecules are 
in general subject to two kinds of reactions: the formation of trimers via
\begin{subequations}
\label{trimer}
\begin{align}
{\rm KRb}(n_1) + {\rm KRb}(n_2) &\rightarrow {\rm K}_2{\rm Rb} + {\rm Rb}
 + \Delta E _{\rm trimer} \label{trimer:1} \\
{\rm KRb}(n_1) + {\rm KRb}(n_2) &\rightarrow {\rm Rb_2K} + {\rm K}
 + \Delta E_{\rm trimer} \label{trimer:2}
\end{align}
\end{subequations}
and the bond-swapping reaction
\begin{eqnarray}
\label{bond_swap}
{\rm KRb}(n_1) + {\rm KRb}(n_2) \rightarrow  {\rm K}_2(n_1^{\prime}) + 
{\rm Rb}_2(n_2^{\prime}) +\Delta E_{\rm bs}.
\end{eqnarray}

We report here two circumstances: 1) The reactions (\ref{trimer})
that form trimers are energetically disallowed at ultracold
temperatures, as the energy released, $\Delta E_{\rm trimer}$, is
negative and large -- several thousand wave numbers, nine orders of magnitude 
larger than translational kinetic energies in the gas. 
2) By contrast, the bond-rearrangement reactions in (\ref{bond_swap})
produce very small energy differences,
since the bonds are all covalent and very similar.   In zero field 
the reaction (\ref{bond_swap}) with $n_1=n_2=0$ is exothermic by
$\Delta E_{\rm dimer} = 10.4$ cm$^{-1}$ 
Moreover, the reactants KRb are polar whereas 
the products K$_2$ and Rb$_2$ are not, thus the products can only be polarized at
comparatively high fields.  Therefore, the relative 
energy $\Delta E_{\rm dimer}({\cal E})$ between reactants and products
is, in principle, a function of the applied electric field ${\cal E}$.
Indeed, at fields on the order of several $\times 10^5$ V/cm, 
$\Delta E_{\rm bs}$ vanishes and the
reaction can be turned off altogether.  At fields smaller than this,
high-lying rotational final states can be disallowed,
thereby changing the possible distribution of product states.

We begin with the first point, that trimer formation (\ref{trimer})
is energetically forbidden.  As the
trimer binding energies have not been measured, we must calculate them
from {\it ab initio} methods.  In general, three spin-1/2 alkali
atoms can combine to form a trimer with total spin $S = 1/2$ (doublet
state) or  an excited state with $S = 3/2$ (quartet state).
Several calculations of doublet states have been achieved, for the homonuclear
trimers K$_3$ \cite{pavelK3} and Li$_3$ \cite{Byrd09_IJQC}, 
and for certain molecular
Li$_2$X systems, where X is an alkali-metal atom \cite{pavelLi2}.
Sold\'{a}n and co-workers have examined the two doublet surfaces of K$_3$ at 
$C_{\rm 2v}$ geometries as well as the conical intersection at the
 equilateral triangle geometry described by the 
$D_{\rm 3h}$ group \cite{pavelK3}.

We employ similar computational techniques as in Refs. \cite{pavelK3,pavelLi2}
to compute doublet ground states, within the 
{\sc molpro} suite of molecular structure codes \cite{molpro}. 
We use the effective core potentials and basis sets of the Stuttgart group 
for the K (ECP10MDF) and Rb (ECP28MDF) atoms~\cite{stutbasis}.  In 
addition, to adequately model three-body forces, we augment these basis
functions with a diffuse function for each of the $s$, 
$p$, and $d$ orbitals in an even tempered manner, as
well as a $g$-function for the K atom.

We compare two computational approaches, testing them on diatomic alkali
molecules. For both approaches, we first  perform a spin-restricted 
Hartree-Fock (RHF) calculation on the singlet configuration. 
The first approach uses the RHF wave function as the foundation for
a multi-configuration self consistent field (MCSCF)
calculation~\cite{mcscf1,mcscf2}
with an active space that includes the first excited $p$ orbital of each atom. 
In addition, we include all states resulting from the $ns^0\,np^1$ 
configurations of each atom.  All
remaining orbitals are closed, meaning they are energy optimized with the 
restriction that they remain doubly occupied.
After each MCSCF calculation, we perform an internally contracted multi-reference 
configuration interaction (MRCI)~\cite{mrci1,mrci2} calculation with the same active space as 
in the MCSCF. The minimum energy is then compared to the dissociation energy
evaluated in the  separated-atom limit.

In the second approach, we perform a 
coupled clusters with single, double and non-iterative triples (CCSD(T))~\cite{ccsd1} 
on the $X{}^1\Sigma$ state at the minimum $R_{\rm e}$ value obtained from the 
MCSCF+MRCI calculation. We then performed a geometry optimization followed by a 
basis set superposition error correction (BSSE)\cite{Boys70_MP}
 to extract the bond length and dissociation 
energy of the systems at the RHF-CCSD(T)+BSSE level of theory.

\begin{table}
  \caption{\label{t:Di-prop} Molecular properties of the diatomic molecules K$_2$, KRb, 
    and Rb$_2$. Equilibrium bond lengths $R_e$ are in $\AA$ and dissociation energies 
    $D_e$ are in 
    cm$^{\rm -1}$. (1) denotes a calculation performed at the MCSCF+MRCI level while 
    (2) denotes a geometry optimization at the RHF-CCSD(T) level accounting for BSSE. }
  \begin{ruledtabular}
    \begin{tabular}{l|l|l|l|l|l|l}
      Molecule & $R_e$ (1) & $D_e$ (1) & $R_e$ (2) & $D_e$ (2) & $R_e$ Expt. 
      & $D_e$ Expt.\\
      \hline
      K$_2$ & 4.16 & -4293 & 3.92 & -4328 & 3.92 & -4450.78(15) \\
      \hline
      KRb & 4.33 & -4039 & 4.05 & -4062 & 4.05 & -4217.30(15)\\
      \hline
      Rb$_2$ & 4.50 & -3729 & 4.18 & -3741 & 4.17 & -3993.53(6)\\
      \hline
    \end{tabular}
  \end{ruledtabular}
\end{table}

To check the adequacy of these computational methods, we compare their results with
the known binding energies of alkali-metal dimers \cite{k2,krb,rb2,rbcs,cs2},
as shown in Table~\ref{t:Di-prop}.  These results show that
approach 1 gets close to the dissociation energy of each diatomic 
species, but noticeably overestimates the bond length, especially for the 
heavier alkali-metal systems. By contrast, approach 2 does just as well at
predicting dissociation energies, but provides far better bond lengths.  In either method,
binding energies are clearly reproduced to within several hundred cm$^{-1}$.
We take this as an empirical measure of the calculations' accuracy. 

We next apply the same methods to determine the minima of the 
three-body potential energy surfaces. 
The minimum of the potential energy surface should lie in the $C_{\rm 2v}$ 
point group i.e., the odd-atom-out is expected to lie on the perpendicular bisector
of the line joining the two like atoms.  Thus we initially 
restrict ourselves to this geometry. As a case 
study, we will look in detail at the K$_2$Rb surfaces, but the Rb$_2$K
surfaces are similar.

As a first approach we used the RHF-MCSCF+MRCI method due to its time efficiency 
To be consistent with the 
method employed for dimers, we keep the same active space, which included the first 
excited $p$-orbital of each atom. We included 4 states each of ${}^2$B$_2$ and 
${}^2$A$_1$ symmetries, 2 ${}^2$B$_1$, 1 ${}^2$A$_2$, and 1 ${}^4$B$_2$ states in 
the MCSCF. 
In each MRCI calculation of the ${}^2$B$_2$ and ${}^2$A$_1$ surfaces, we 
included 4 states as well as the reference symmetries of the other 
doublets. 

\begin{figure}
\label{fig1}
  \begin{center}
    \resizebox{3.5in}{!}{\includegraphics{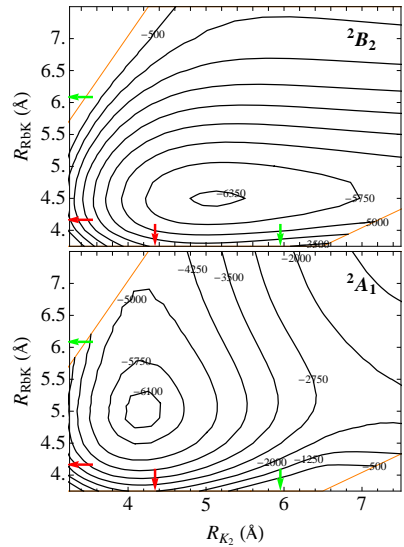}} 
  \end{center} 
  \caption{(Color Online) Contour plots for the doublet ground state PES's of
    K$_2$Rb, versus interatomic spacings in an isosceles triangle geometry.
    The top panel shows the ${}^2$B$_2$ surface while the bottom panel shows
    the ${}^2$A$_1$ surface.  Contours are labeled in 
    increments of 750~cm$^{-1}$ from -5750 -- -500 with an additional contour 
    near the minimum of each surface.  Arrows denote the singlet (red) and
    triplet (green) equilibrium bond lengths of the diatomic species.} 
\end{figure}

The results of the RHF-MCSCF+MRCI calculations are presented in 
Fig. 1 as contour plots of the potential energy surface (PES) 
in the two independent bond lengths
$R_{\rm RbK}$ and $R_{\rm K_2}$, for the isosceles triangle geometry.
These surfaces show that the $^2B_2$ surface (upper panel) possesses the global minimum,
hence represents the ground state of the trimer.  The minimum energy occurs
near the singlet bond length of the KRb dimer (denoted by a horizontal
red arrow), but is intermediate between the singlet and 
triplet bond lengths of the K$_2$ dimer (red and green vertical arrows, respectively).
In C$_{\rm 2v}$ symmetries, B$_2$ corresponds to an odd reflection 
about the bisector of the isosceles triangle. This means that along the K$_2$ 
bond the electronic wave function should roughly resemble the K$_2$ triplet
wave function, and so should the the bond length.  There is then diminished  electron 
density at the center between the two K atoms, and this 
allows the Rb atom to fill this space,  bringing the two K atoms slightly 
closer together than they would be in the K$_2$ triplet state alone. 

Similarly, the minimum of the ${}^2$A$_1$ surface in the lower panel of Fig.~1
is located near the K$_2$ singlet bond length (vertical red arrow),
but in between the singlet and triplet bonds 
of the KRb system (horizontal red and green arrows). The $^2$A$_1$ surface  
requires an even reflection in the electronic wave function across the 
bisector of the isosceles triangle, therefore preferring a singlet-like
 bond in $R_{\rm K_2}$. Therefore, the Rb atom does not 
quite know which spin to take since it can form a triplet or singlet with one 
or the other K atoms, but not both. This frustration manifests itself 
with a bond somewhere intermediate between singlet and triplet bonds in
this coordinate.

Knowing that the the RHF-CCSD(T)+BSSE results in the dimer case are markedly 
better when compared to experimental values, we choose to  
characterize the minima of the PES's at RHF-UCCSD(T)+BSSE level of theory,
including a basis set superposition correction. The U in UCCSD(T) 
refers to the spin-unrestricted formalism due to the open-shell nature of the 
trimer systems.  Starting near the parameters obtained above for the 
${}^2$B$_2$ surface, we find the optimum geometry by the method of steepest descents. 
Binding energies are determined at the RHF-UCCSD(T) level of theory, accounting for basis 
set overlap errors. The results are presented in Table~\ref{t:Tri-prop} for all the 
alkali-metal trimers containing K and Rb. In the case of
K$_3$, where results have previously been computed, we find
excellent agreement with the reported values 
of the isosceles bond length $r_{\rm iso}=4.10$~$\AA$ and apex angle 
 $\theta_{\rm apex}=77.13^\circ$~\cite{pavelK3}.  Based on these results, we
conclude that the formation of the K$_2$Rb trimer would require
$\Delta E_{\rm trimer} = 2(4217)-5982$ $=2452$ cm$^{-1}$ of energy,
certainly rendering this reaction impossible at ultralow temperatures.

While a minimum of the surface appears to be in the B$_2$ symmetry of 
the C$_{\rm 2v}$ point group, an examination of the trimers K$_2$Rb and Rb$_2$K 
away from this symmetry is needed in order to determine whether this is a global minumum. 
We looked at the systems in the C$_{\rm s}$ point group and found minimums which correspond 
to bent geometries using similar optimaization procedures. In both trimers the bent geomtry 
minimum is found to be the global minimum making the B$_2$ minimum a saddle point. The 
parameters are given in Table~\ref{t:Tri-prop}. $r_{\rm hetero}$ refers to the shorter of 
the two bonds between heteronuclear pairs when in C$_{\rm s}$ symmetry. $\theta$ is the 
angle made by A-q-B, where A and B are differing atoms and q is the midpoint between the 
homonuclear bond.

\begin{table} 
  \caption{\label{t:Tri-prop} Molecular properties of the alkali-metal triatomic molecules 
    containing K and Rb. Bond lengths are in $\AA$, angles are in degrees, 
    and dissociation energies, relative to the complete breakup
    into three free atoms, are in cm$^{\rm -1}$. }
  \begin{ruledtabular}
    \begin{tabular}{l|l|l|l|l|l}
      Molecule & Symmetry & $r_{\rm hetero}$ & $r_{\rm homo}$ & $\theta$ & $D_e$\\
      \hline
      K$_3$ & C$_{\rm 2v}$ $^2$B$_2$ & 4.10 & 5.11 & 90.0 & -6059\\
      \hline
      K$_2$Rb & C$_{\rm 2v}$ $^2$B$_2$ & 4.26 & 4.99 & 90.0 & -5817\\
      \hline
      Rb$_2$K & C$_{\rm 2v}$ $^2$B$_2$ & 4.26 & 5.63 & 90.0 & -5533\\
      \hline
      K$_2$Rb & C$_{\rm s}$ $^2$A$^\prime$ & 4.25 & 4.06 & 73.23 & -5982\\
      \hline
      Rb$_2$K & C$_{\rm s}$ $^2$A$^\prime$ & 4.24 & 4.41 & 75.00 & -5611\\
      \hline
      Rb$_3$ & C$_{\rm 2v}$ $^2$B$_2$ & 4.40 & 5.59 & 90.0 & -5273\\
      \hline
    \end{tabular}
  \end{ruledtabular}
\end{table}

\begin{figure}\label{f:krbPkrb}
  \begin{center}
    \resizebox{3.75in}{!}{\includegraphics{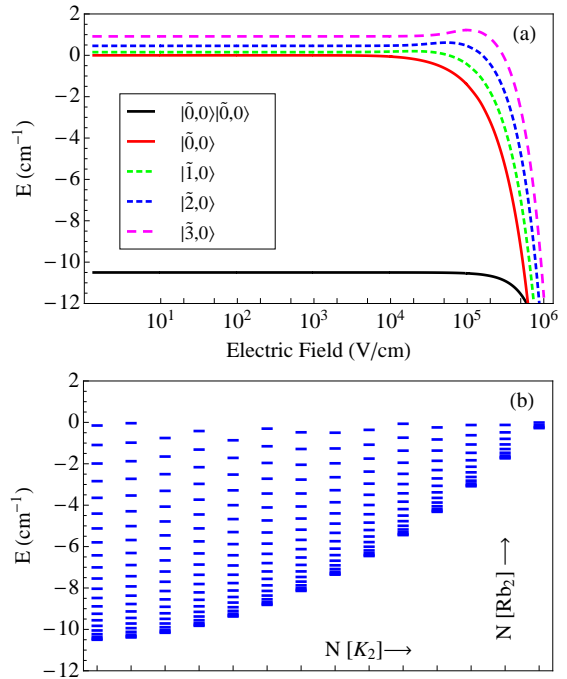}} 
  \end{center} \vspace{-.2in}
  \caption{(Color Online) Panel (a) shows the Stark shift for the ground and first few 
    rotationally excited states for  a pair of KRb molecules (colored lines). The
    black line shows the energy of the final states K$_2(n_1^{\prime}=0)$ + 
    Rb$_2(n_2^{\prime}=0)$.  Panel (b) shows the rotational states energetically allowed
    at zero electric field.  
    }
\end{figure}

We therefore
now consider the bond-swapping reaction (\ref{bond_swap}), and in particular
the variation of $\Delta E_{\rm bs}$ with electric field. To do so, we look at the
energy eigenvalues of the reactant and product molecules, as given by the Hamiltonian
\begin{equation}\label{e:hamiltonian}
  H = B_e{\vec N}^2 - D~{\vec N}^4-\vec{d}\cdot\vec{\mathcal{E}}
\end{equation}
where $B_e$ is the molecule's rotational constant, $D$ is centrifugal distortion
constant, ${\vec d}$ is the body-fixed molecular dipole moment (which is zero for the
products), and ${\vec {\cal E}}$ is the applied electric field.  In addition, both
kinds of molecules also experience a shift due to their electronic polarizability $\alpha$.
Using the known KRb dipole moment ~\cite{Ni08_Science} and computed molecular
polarizabilities \cite{polarizabilities}, we can
compute the relative energies of the reactants and products.

In Fig.2a) we plot the energy of the ground state (red) and several
rotationally excited states of KRb+KRb as a function of electric field.  Also
shown (black) is the ground state energy of the products, K$_2(n=0$) + Rb$_2(n=0)$.
In zero field, the reactants are their natural 10.4 cm$^{-1}$ above the products.
As the field increases, the energy of the polar reactant states decreases rapidly,
but that of the non-polar product states decreases far more slowly.  Therefore,
at a field beyond $\sim 625$ kV/cm, the reactants are actually lower in energy
than the products, $\Delta E_{\rm bs}$ becomes negative,
and the reaction is completely shut off.  While a static electric field this
large is probably impractical to implement, it may be possible to achieve
the required energy shifts in a suitably designed microwave cavity, such as
those proposed for trapping polar species \cite{Demille04_EPJD}.  Theory would then
naturally have to account for collisions of the field-dressed 
states~\cite{Avdeenkov09_NJP,Alyabyshev09_PRA}.

As a point of reference, Fig. 2b) shows the 235 energetically allowed 
states of the products in zero electric field, indexed by their rotational 
quantum numbers $n_1^{\prime}$ and $n_2^{\prime}$.  
It is clear that, before the electric field shuts off
all reactions completely, it shuts off first higher-$n^{\prime}$, then successively
lower-$n^{\prime}$ states.  Recall that the $n^{\prime}$-distribution of 
the products is one of the key
observables of physical chemistry.  The ability to allow only certain values 
of $n^{\prime}$ into this distribution will likely provide an even greater wealth of
information from such experiments.  The detailed effect on chemistry 
of shutting off 
successively lower rotational exit channels remains to be explored.
It may be hoped, for example, that the new information gleaned would
shed additional light on the role in these reactions of conical intersections.

So far we have focused on KRb, since it is the molecule for which 
ultracold chemistry has recently been demonstrated experimentally.  
However, there are other reasonable candidates for these experiments,
notably RbCs~\cite{rbcspaper}, whose reaction is {\it endothermic} by 28.7 cm$^{-1}$.
Such a reaction would proceed only by placing the reactant RbCs molecules
into excited states, which could certainly be done. An applied
electric field could then still dictate which final channels 
are available, by moving these rotationally excited states
relative to the Rb$_2$ $+$ Cs$_2$ products.

{\it Note added in preparation:} New work shows that the timer reactions
(\ref{trimer}) are energetically forbidden for all alkali dimer collisions
\cite{Zuchowski10_preprint}.

The authors acknowledge useful discussions with G. Qu\'{e}m\'{e}ner and
C. H. Greene, as well as funding from the NSF.

\end{document}